\documentclass[a4paper]{article}

\usepackage{INTERSPEECH2021}
\usepackage{color}
\usepackage{url}
\usepackage{booktabs}

\definecolor{dblue}{RGB}{0,0,255}

\newcommand{\squeezeup}{\vspace{-2.0mm}}
\newcommand\blfootnote[1]{%
  \begingroup
  \renewcommand\thefootnote{}\footnote{#1}%
  \addtocounter{footnote}{-1}%
  \endgroup
}

\title{AST: Audio Spectrogram Transformer}
\name{Yuan Gong, Yu-An Chung, James Glass}
\address{MIT Computer Science and Artificial Intelligence Laboratory, Cambridge, MA 02139, USA}
\email{\{yuangong, andyyuan, glass\}@mit.edu}

\begin{document}

\maketitle
\begin{abstract}

In the past decade, convolutional neural networks~(CNNs) have been widely adopted as the main building block for end-to-end audio classification models, which aim to learn a direct mapping from audio spectrograms to corresponding labels. To better capture long-range global context, a recent trend is to add a self-attention mechanism on top of the CNN, forming a CNN-attention hybrid model. However, it is unclear whether the reliance on a CNN is necessary, and if neural networks purely based on attention are sufficient to obtain good performance in audio classification. In this paper, we answer the question by introducing the \emph{Audio Spectrogram Transformer}~(AST), the first convolution-free, purely attention-based model for audio classification. We evaluate AST on various audio classification benchmarks, where it achieves new state-of-the-art results of~0.485 mAP on AudioSet,~95.6\% accuracy on ESC-50, and~98.1\% accuracy on Speech Commands V2. 

\end{abstract}

\section{Introduction}
\label{sec:intro}

With the advent of deep neural networks, over the last decade audio classification research has moved from models based on hand-crafted features~\cite{eyben2013recent,schuller2013interspeech} to end-to-end models that directly map audio spectrograms to corresponding labels~\cite{jaitly2011learning,dieleman2014end,trigeorgis2016adieu}.
Specifically, convolutional neural networks~(CNNs)~\cite{lecun1995convolutional} have been widely used to learn representations from raw spectrograms for end-to-end modeling, as the inductive biases inherent to CNNs such as spatial locality and translation equivariance are believed to be helpful.
In order to better capture long-range global context, a recent trend is to add a self-attention mechanism on top of the CNN.
Such CNN-attention hybrid models have achieved state-of-the-art~(SOTA) results for many audio classification tasks such as audio event classification~\cite{kong2020panns,gong2021psla}, speech command recognition~\cite{rybakov2020streaming}, and emotion recognition~\cite{li2018attention}.
However, motivated by the success of purely attention-based models in the vision domain~\cite{dosovitskiy2021image,touvron2020deit,yuan2021tokens}, it is reasonable to ask whether a CNN is still essential for audio classification. 
\blfootnote{Code at \url{https://github.com/YuanGongND/ast}.}

To answer the question, we introduce the \emph{Audio Spectrogram Transformer}~(AST), a \emph{convolution-free}, \emph{purely} attention-based model that is directly applied to an audio spectrogram and can capture long-range global context even in the lowest layers. 
Additionally, we propose an approach for transferring knowledge from the Vision Transformer~(ViT)~\cite{touvron2020deit} pretrained on ImageNet~\cite{deng2009imagenet} to AST, which can significantly improve the performance.
The advantages of AST are threefold.
First, AST has superior performance: we evaluate AST on a variety of audio classification tasks and datasets including AudioSet~\cite{gemmeke2017audio}, ESC-50~\cite{piczak2015esc} and Speech Commands~\cite{warden2018speech}.
AST outperforms state-of-the-art systems on all these datasets.
Second, AST naturally supports variable-length inputs and can be applied to different tasks without any change of architecture.
Specifically, the models we use for all aforementioned tasks have the same architecture while the input lengths vary from~1 sec.~(Speech Commands) to~10 sec.~(AudioSet).
In contrast, CNN-based models typically require architecture tuning to obtain optimal performance for different tasks.
Third, comparing with SOTA CNN-attention hybrid models, AST features a simpler architecture with fewer parameters, and converges faster during training.
To the best of our knowledge, AST is the first purely attention-based audio classification model.

\begin{figure}[t]
  \centering
  \includegraphics[width=7.0cm]{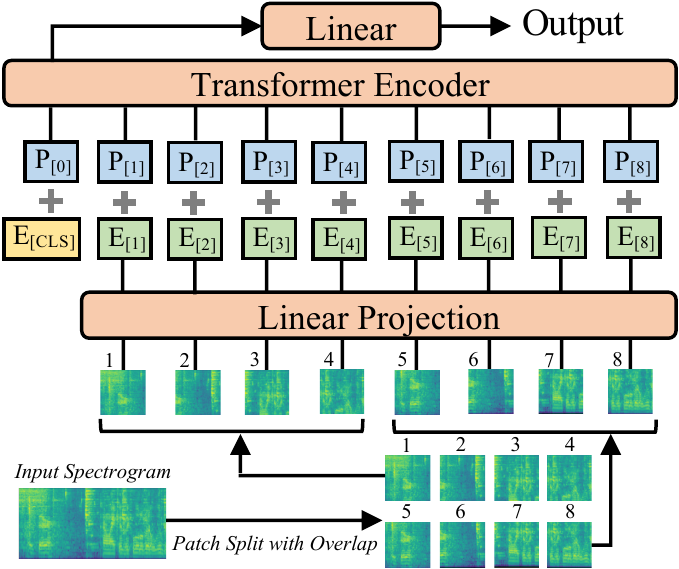}
  \caption{The proposed audio spectrogram transformer (AST) architecture. The 2D audio spectrogram is split into a sequence of 16$\times$16 patches with overlap, and then linearly projected to a sequence of 1-D patch embeddings. Each patch embedding is added with a learnable positional embedding. An additional classification token is prepended to the sequence. The output embedding is input to a Transformer, and the output of the classification token is used for classification with a linear layer.}
  \label{fig:arc}
  \squeezeup \squeezeup
\end{figure}

\textbf{Related Work} The proposed Audio Spectrogram Transformer, as the name suggests, is based on the Transformer architecture~\cite{vaswani2017attention}, which was originally proposed for natural language processing tasks. Recently, the Transformer has also been adapted for audio processing, but is typically used in conjunction with a CNN~\cite{miyazaki2020convolution,kong2020sound,gulati2020conformer}. In~\cite{miyazaki2020convolution,kong2020sound}, the authors stack a Transformer on top of a CNN, while in~\cite{gulati2020conformer}, the authors combine a Transformer and a CNN in each model block. Other efforts combine CNNs with simpler attention modules~\cite{gong2021psla,kong2020panns,rybakov2020streaming}. The proposed AST differs from these studies in that it is convolution-free and purely based on attention mechanisms. The closest work to ours is the Vision Transformer~(ViT)~\cite{dosovitskiy2021image,touvron2020deit,yuan2021tokens}, which is a Transformer architecture for vision tasks. AST and ViT have similar architectures but ViT has only been applied to fixed-dimensional inputs~(images) while AST can process variable-length audio inputs. In addition, we propose an approach to transfer knowledge from ImageNet pretrained ViT to AST. We also conduct extensive experiments to show the design choice of AST on audio tasks.

\section{Audio Spectrogram Transformer}
\subsection{Model Architecture}
\label{sec:arc}

Figure~\ref{fig:arc} illustrates the proposed Audio Spectrogram Transformer~(AST) architecture.
First, the input audio waveform of~$t$ seconds is converted into a sequence of~128-dimensional log Mel filterbank~(fbank) features computed with a~25ms Hamming window every~10ms.
This results in a~$128\times100t$ spectrogram as input to the AST.
We then split the spectrogram into a sequence of~$N$ $16\times16$ patches with an overlap of~6 in both time and frequency dimension, where $N=12\lceil{(100t-16)/10}\rceil$ is the number of patches and the effective input sequence length for the Transformer.
We flatten each~$16\times16$ patch to a 1D patch embedding of size~768 using a linear projection layer.
We refer to this linear projection layer as the patch embedding layer.
Since the Transformer architecture does not capture the input order information and the patch sequence is also not in temporal order, we add a trainable positional embedding~(also of size~768) to each patch embedding to allow the model to capture the spatial structure of the 2D audio spectrogram.

Similar to~\cite{devlin2019bert}, we append a \texttt{[CLS]} token at the beginning of the sequence.
The resulting sequence is then input to the Transformer.
A Transformer consists of several encoder and decoder layers.
Since AST is designed for classification tasks, we only use the encoder of the Transformer.
Intentionally, we use the original Transformer encoder~\cite{vaswani2017attention} architecture without modification.
The advantages of this simple setup are 1) the standard Transformer architecture is easy to implement and reproduce as it is off-the-shelf in TensorFlow and PyTorch, and 2) we intend to apply transfer learning for AST, and a standard architecture makes transfer learning easier.
Specifically, the Transformer encoder we use has an embedding dimension of~768, 12 layers, and 12 heads, which are the same as those in~\cite{touvron2020deit,dosovitskiy2021image}.
The Transformer encoder's output of the \texttt{[CLS]} token serves as the audio spectrogram representation.
A linear layer with sigmoid activation maps the audio spectrogram representation to labels for classification. 

Strictly speaking, the patch embedding layer can be viewed as a single convolution layer with a large kernel and stride size, and the projection layer in each Transformer block is equivalent to 1$\times$1 convolution. However, the design is different from conventional CNNs that have multiple layers and small kernel and stride sizes. These Transformer models are usually referred to as convolution-free to distinguish them from CNNs ~\cite{dosovitskiy2021image,touvron2020deit}.

\subsection{ImageNet Pretraining}
\label{sec:pretrain}

One disadvantage of the Transformer compared with CNNs is that the Transformer needs more data to train~\cite{dosovitskiy2021image}. In~\cite{dosovitskiy2021image}, the authors point out that the Transformer only starts to outperform CNNs when the amount of data is over~14 million for image classification tasks. However, audio datasets typically do not have such large amounts of data, which motivates us to apply cross-modality transfer learning to AST since images and audio spectrograms have similar formats. Transfer learning from vision tasks to audio tasks has been previously studied in~\cite{gwardys2014deep,guzhov2020esresnet,palanisamy2020rethinking,gong2021psla}, but only for CNN-based models, where ImageNet-pretrained CNN weights are used as initial CNN weights for audio classification training. In practice, it is computationally expensive to train a state-of-the-art vision model, but many commonly used architectures (e.g., ResNet~\cite{he2016deep}, EfficientNet~\cite{tan2019efficientnet}) have off-the-shelf ImageNet-pretrained models for both TensorFlow and PyTorch, making transfer learning much easier. We also follow this regime by adapting an off-the-shelf pretrained Vision Transformer (ViT) to AST. 


While ViT and AST have similar architectures (e.g., both use a standard Transformer, same patch size, same embedding size), they are not same. Therefore, a few modifications need to make for the adaptation. First, the input of ViT is a 3-channel image while the input to the AST is a single-channel spectrogram, we average the weights corresponding to each of the three input channels of the ViT patch embedding layer and use them as the weights of the AST patch embedding layer. This is equivalent to expanding a single-channel spectrogram to 3-channels with the same content, but is computationally more efficient. We also normalize the input audio spectrogram so that the dataset mean and standard deviation are 0 and 0.5, respectively. Second, the input shape of ViT is fixed (either $224\times224$ or $384\times384$), which is different from a typical audio spectrogram. In addition, the length of an audio spectrogram can be variable. While the Transformer naturally supports variable input length and can be directly transferred from ViT to AST, the positional embedding needs to be carefully processed because it learns to encode the spatial information during the ImageNet training. We propose a cut and bi-linear interpolate method for positional embedding adaptation. For example, for a ViT that takes $384\times384$ image input and uses a patch size of $16\times16$, the number of patches and corresponding positional embedding is $24\times24=576$ (ViT splits patches without overlap). An AST that takes 10-second audio input has $12\times100$ patches, each patch needs a positional embedding. We therefore cut the first dimension and interpolate the second dimension of the $24\times24$ ViT positional embedding to $12\times100$ and use it as the positional embedding for the AST. We directly reuse the positional embedding for the \texttt{[CLS]} token. By doing this we are able to transfer the 2D spatial knowledge from a pretrained ViT to the AST even when the input shapes are different. Finally, since the classification task is essentially different, we abandon the last classification layer of the ViT and reinitialize a new one for AST. With this adaptation framework, the AST can use various pretrained ViT weights for initialization. In this work, we use pretrained weights of a data-efficient image Transformer~(DeiT)~\cite{touvron2020deit}, which is trained with CNN knowledge distillation, $384\times384$ images, has 87M parameters, and achieves 85.2\% top-1 accuracy on ImageNet 2012. During ImageNet training, DeiT has two \texttt{[CLS]} tokens; we average them as a single \texttt{[CLS]} token for audio training.

\section{Experiments}

In this section, we focus on evaluating the AST on AudioSet (Section~\ref{sec:audioset}) as weakly-labeled audio event classification is one of the most challenging audio classification tasks. We present our primary AudioSet results and ablation study in Section~\ref{sec:audiores} and Section~\ref{sec:ablation}, respectively. We then present our experiments on ESC-50 and Speech Commands V2 in Section~\ref{sec:transfer}.

\subsection{AudioSet Experiments}
\label{sec:audioset}
\subsubsection{Dataset and Training Details}
AudioSet~\cite{gemmeke2017audio} is a collection of over 2 million 10-second audio clips excised from YouTube videos and labeled with the sounds that the clip contains from a set of 527 labels. The balanced training, full training, and evaluation set contains 22k, 2M, and 20k samples, respectively. For AudioSet experiments, we use the exact same training pipeline with~\cite{gong2021psla}. Specifically, we use ImageNet pretraining (as described in Section~\ref{sec:pretrain}), balanced sampling (for full set experiments only), data augmentation (including mixup~\cite{tokozume2018learning} with mixup ratio$=$0.5 and spectrogram masking~\cite{park2019specaugment} with max time mask length of 192 frames and max frequency mask length of 48 bins), and model aggregation (including weight averaging~\cite{izmailov2018averaging} and ensemble~\cite{breiman1996bagging}). We train the model with a batch size of 12, the Adam optimizer~\cite{kingma2015adam}, and use binary cross-entropy loss. We conduct experiments on the official balanced and full training set and evaluate on the AudioSet evaluation set. For balanced set experiments, we use an initial learning rate of 5e-5 and train the model for 25 epochs, the learning rate is cut into half every 5 epoch after the 10th epoch. For full set experiments, we use an initial learning rate of 1e-5 and train the model for 5 epochs, the learning rate is cut into half every epoch after the 2nd epoch. We use the mean average precision (mAP) as our main evaluation metric.

\begin{table}[]
\footnotesize
\centering
\caption{Performance comparison of AST and previous methods on AudioSet.}
\label{tab:audiosetexp}
\begin{tabular}{@{}llcc@{}}
\toprule
         & \multicolumn{1}{c}{\begin{tabular}[c]{@{}c@{}}Model \\ Architecture\end{tabular}} & \multicolumn{1}{c}{\begin{tabular}[c]{@{}c@{}}Balanced \\ mAP\end{tabular}} & \multicolumn{1}{c}{\begin{tabular}[c]{@{}c@{}}Full \\ mAP\end{tabular}} \\ \midrule
Baseline~\cite{gemmeke2017audio} & CNN+MLP               &             -                                                                    &        0.314                                                                    \\

PANNs~\cite{kong2020panns}    & CNN+Attention                &       0.278                                                                          &   0.439                                                                         \\
PSLA~\cite{gong2021psla} (Single)     & CNN+Attention             &  0.319                                                                               &    0.444                                                                        \\ 
PSLA (Ensemble-S)     & CNN+Attention                 &   0.345                                                                              &   0.464                                                                         \\
PSLA (Ensemble-M)     & CNN+Attention                 &   0.362                                                                              &   0.474                                                                         \\ \midrule
AST (Single)      & Pure Attention                 &  \begin{tabular}[c]{@{}c@{}}0.347\\ $\pm$ 0.001\end{tabular}                                                                              &  \begin{tabular}[c]{@{}c@{}}0.459\\ $\pm$ 0.000\end{tabular}
\\
AST (Ensemble-S)      & Pure Attention                 &  0.363                                                                               &  0.475
\\
AST (Ensemble-M)      & Pure Attention                 &  \bf{0.378}                                                                               &  \bf{0.485}
\\ \bottomrule
\end{tabular}
\squeezeup
\end{table}

\subsubsection{AudioSet Results}
\label{sec:audiores}

We repeat each experiment three times with the same setup but different random seeds and report the mean and standard deviation. When AST is trained with the full AudioSet, the mAP at the last epoch is 0.448$\pm$0.001. As in~\cite{gong2021psla}, we also use weight averaging~\cite{izmailov2018averaging} and ensemble~\cite{breiman1996bagging} strategies to further improve the performance of AST. Specifically, for weight averaging, we average all weights of the model checkpoints from the first to the last epoch. The weight-averaged model achieves an mAP of 0.459$\pm$0.000, which is our best single model (weight averaging does not increase the model size). For ensemble, we evaluate two settings: 1) Ensemble-S: we run the experiment three times with the exact same setting, but with a different random seed. We then average the output of the last checkpoint model of each run. In this setting, the ensemble model achieves an mAP of 0.475; 2) Ensemble-M: we ensemble models trained with different settings, specifically, we ensemble the three models in Ensemble-S together with another three models trained with different patch split strategies (described in Section~\ref{sec:ablation} and shown in Table~\ref{tab:overlap}). In this setting, the ensemble model achieves an mAP of 0.485, this is our best full model on AudioSet. As shown in Table~\ref{tab:audiosetexp}, the proposed AST outperforms the previous best system in~\cite{gong2021psla} in all settings. Note that we use the same training pipeline with~\cite{gong2021psla} and~\cite{gong2021psla} also use ImageNet pretraining, so it is a fair comparison. In addition, we use fewer models (6) for our best ensemble models than~\cite{gong2021psla} (10). Finally, it is worth mentioning that AST training converges quickly; AST only needs 5 training epochs, while in~\cite{gong2021psla}, the CNN-attention hybrid model is trained for 30 epochs.

We also conduct experiments with the balanced AudioSet (about $1\%$ of the full set) to evaluate the performance of AST when the training data volume is smaller. For weight averaging, we average all weights of the model checkpoints of the last 20 epochs. For Ensemble-S, we follow the same setting used for the full AudioSet experiment; for Ensemble-M, we include 11 models trained with different random seeds (Table~\ref{tab:audiosetexp}), different pretrained weights (Table~\ref{tab:arc}), different positional embedding interpolation (Table~\ref{tab:position}), and different patch split strategies (Table~\ref{tab:overlap}). The single, Ensemble-S, and Ensemble-M model achieve 0.347$\pm$0.001, 0.363, and 0.378, respectively, all outperform the previous best system. This demonstrates that AST can work better than CNN-attention hybrid models even when the training set is relatively small.

\subsubsection{Ablation Study}
\label{sec:ablation}

We conduct a series of ablation studies to illustrate the design choices for the AST. To save compute, we mainly conduct ablation studies with the balanced AudioSet. For all experiments, we use weight averaging but do not use ensembles.

\begin{table}[t]
\footnotesize
\centering
\caption{Performance impact due to ImageNet pretraining. ``Used'' denotes the setting used by our optimal AST model.}
\label{tab:pretrain}
\begin{tabular}{@{}lccc@{}}
\toprule
                               & Balanced Set & Full Set  \\ \midrule
No Pretrain                  &  0.148           &   0.366                  \\
ImageNet Pretrain (Used)     &  0.347        &   0.459                  \\ \bottomrule
\end{tabular}
\end{table}

\begin{table}[t]
\footnotesize
\centering
\caption{Performance of AST models initialized with different ViT weights on balanced AudioSet and corresponding ViT models' top-1 accuracy on ImageNet 2012. (* Model is trained without patch split overlap due to memory limitation.)}
\label{tab:arc}
\begin{tabular}{@{}lcccc@{}}
\toprule
                               & \# Params & ImageNet & AudioSet  \\ \midrule
ViT Base~\cite{dosovitskiy2021image}           &  86M      &0.846      &   0.320                  \\
ViT Large~\cite{dosovitskiy2021image}*         &  307M     &0.851       &   0.330                  \\
DeiT w/o Distill~\cite{touvron2020deit}          &  86M      &0.829      &   0.330                  \\
DeiT w/ Distill (Used)        &  87M     &0.852       &   0.347                  \\
\bottomrule
\end{tabular}
\squeezeup
\end{table}

\newcommand{\myparagraph}[1]{\vspace{.4em} \noindent \textbf{#1}\ }
\myparagraph{Impact of ImageNet Pretraining.} We compare ImageNet pretrained AST and randomly initialized AST. As shown in Table~\ref{tab:pretrain}, ImageNet pretrained AST noticeably outperforms randomly initialized AST for both balanced and full AudioSet experiments. The performance improvement of ImageNet pretraining is more significant when the training data volume is smaller, demonstrating that ImageNet pretraining can greatly reduce the demand for in-domain audio data for AST. We further study the impact of pretrained weights used. As shown in Table~\ref{tab:arc}, we compare the performance of AST models initialized with pretrained weights of ViT-Base, ViT-Large, and DeiT models. These models have similar architectures but are trained with different settings. We made the necessary architecture modifications for AST to reuse the weights. We find that AST using the weights of the DeiT model with distillation that performs best on ImageNet2012 also performs best on AudioSet. 

\begin{table}[t]
\footnotesize
\centering
\caption{Performance impact due to various positional embedding adaptation settings.}
\label{tab:position}
\begin{tabular}{@{}lcc@{}}
\toprule
                                & Balanced Set  \\ \midrule
Reinitialize                         &     0.305                \\
Nearest Neighbor Interpolation                     &     0.346                \\
Bilinear Interpolation (Used)                      &     0.347                \\ \bottomrule
\end{tabular}
\end{table}

\myparagraph{Impact of Positional Embedding Adaptation.} As mentioned in Section~\ref{sec:pretrain}, we use a cut and bi-linear interpolation approach for positional embedding adaptation when transferring knowledge from the Vision Transformer to the AST. We compare it with a pretrained AST model with a randomly initialized positional embedding. As shown in Table~\ref{tab:position}, we find reinitializing the positional embedding does not completely break the pretrained model as the model still performs better than a fully randomly reinitialized model, but it does lead to a noticeable performance drop compared with the proposed adaptation approach. This demonstrates the importance of transferring spatial knowledge. Bi-linear interpolation and nearest-neighbor interpolation do not result in a big difference.

\begin{table}[t]
\footnotesize
\centering
\caption{Performance impact due to various patch overlap size.}
\label{tab:overlap}
\begin{tabular}{@{}lccc@{}}
\toprule
                               & \# Patches & Balanced Set & Full Set \\ \midrule
No Overlap                     &  512            &    0.336  & 0.451              \\
Overlap-2                      &  657            &    0.342  & 0.456              \\
Overlap-4                      &  850            &    0.344  & 0.455           \\
Overlap-6 (Used)               &  1212            &   0.347 &  0.459              \\\bottomrule
\end{tabular}
\end{table}

\begin{table}[!t]
\footnotesize
\centering
\caption{Performance impact due to various patch shape and size. All models are trained with no patch split overlap.}
\label{tab:patchsize}
\begin{tabular}{@{}lcccc@{}}
\toprule
                               & \# Patches   & w/o Pretrain & w/ Pretrain  \\ \midrule
128$\times$2                   &  512         & 0.154    &   -           \\
16$\times$16 (Used)            &  512         & 0.143       &  0.336            \\ 
32$\times$32                   &  128         & 0.139      &   -           \\
\bottomrule
\end{tabular}
\squeezeup
\end{table}

\myparagraph{Impact of Patch Split Overlap.} We compare the performance of models trained with different patch split overlap~\cite{yuan2021tokens}. As shown in Table~\ref{tab:overlap}, the performance improves with the overlap size for both balanced and full set experiments. However, increasing the overlap also leads to longer patch sequence inputs to the Transformer, which will quadratically increase the computational overhead. Even with no patch split overlap, AST can still outperform the previous best system in~\cite{gong2021psla}.

\myparagraph{Impact of Patch Shape and Size.} As mentioned in Section~\ref{sec:arc}, we split the audio spectrogram into $16\times16$ square patches, so the input sequence to the Transformer cannot be in temporal order. We hope the positional embedding can learn to encode the 2D spatial information. An alternative way to split the patch is slicing the audio spectrogram into rectangular patches in the temporal order. We compare both methods in Table~\ref{tab:patchsize}, when the area of the patch is the same~(256), using $128\times2$ rectangle patches leads to better performance than using $16\times16$ square patches when both models are trained from scratch. However, considering there is no $128\times2$ patch based ImageNet pretrained models, using $16\times16$ patches is still the current optimal solution. We also compare using patches with different sizes, smaller size patches lead to better performance.

\begin{table}[t]
\footnotesize
\centering
\caption{Comparing AST and SOTA models on ESC-50 and Speech Commands. ``-S'' and ``-P'' denotes model trained without and with additional audio data, respectively.}
\label{tab:transfer}
\begin{tabular}{@{}lcccc@{}}
\toprule
                         & ESC-50 & Speech Commands V2 (35 classes)  \\ \midrule
SOTA-S  &          86.5~\cite{sailor2017unsupervised}                  &     97.4~\cite{majumdar2020matchboxnet}                                                                                                  \\
SOTA-P  &        94.7~\cite{kong2020panns}                    &     97.7~\cite{lin2020training}                                                                                      \\ \midrule
AST-S                       &   88.7$\pm$0.7                    &   \bf{98.11$\pm$0.05}                                                                           \\

AST-P                     &  \bf{95.6$\pm$0.4}                         &   97.88$\pm$0.03                                                                           \\
\bottomrule
\end{tabular}
\squeezeup
\end{table}

\subsection{Results on ESC-50 and Speech Commands}
\label{sec:transfer}
The ESC-50~\cite{piczak2015esc} dataset consists of 2,000 5-second environmental audio recordings organized into 50 classes. The current best results on ESC-50 are 86.5\% accuracy (trained from scratch, SOTA-S)~\cite{sailor2017unsupervised} and 94.7\% accuracy (with AudioSet pretraining, SOTA-P)~\cite{kong2020panns}. We compare AST with the SOTA models in these two settings, specifically, we train an AST model with only ImageNet pretraining (AST-S) and an AST model with ImageNet and AudioSet pretraining (AST-P). We train both models with frequency/time masking~\cite{park2019specaugment} data augmentation, a batch size of 48, and the Adam optimizer~\cite{kingma2015adam} for 20 epochs. We use an initial learning rate of 1e-4 and 1e-5 for AST-S and AST-P, respectively, and decrease the learning rate with a factor of 0.85 every epoch after the 5th epoch. We follow the standard 5-fold cross-validation to evaluate our model, repeat each experiment three times, and report the mean and standard deviation. As shown in Table~\ref{tab:transfer}, AST-S achieves 88.7$\pm$0.7 and AST-P achieves 95.6$\pm$0.4, both outperform SOTA models in the same setting. Of note, although ESC-50 has 1,600 training samples for each fold, AST still works well with such a small amount of data even without AudioSet pretraining.


Speech Commands V2~\cite{warden2018speech} is a dataset consists of 105,829 1-second recordings of 35 common speech commands. The training, validation, and test set contains 84,843, 9,981, and 11,005 samples, respectively. We focus on the 35-class classification task, the SOTA model on Speech Commands V2 (35-class classification) without additional audio data pretraining is the time-channel separable convolutional neural network~\cite{majumdar2020matchboxnet}, which achieves~97.4\% on the test set. In~\cite{lin2020training}, a CNN model pretrained with additional 200 million YouTube audio achieves 97.7\% on the test set. We also evaluate AST in these two settings. Specifically, we train an AST model with only ImageNet pretraining (AST-S) and an AST model with ImageNet and AudioSet pretraining (AST-P). We train both models with frequency and time masking~\cite{park2019specaugment}, random noise, and mixup~\cite{tokozume2018learning} augmentation, a batch size of 128, and the Adam optimizer~\cite{kingma2015adam}. We use an initial learning rate of 2.5e-4 and decrease the learning rate with a factor of 0.85 every epoch after the 5th epoch. We train the model for up to 20 epochs, and select the best model using the validation set, and report the accuracy on the test set. We repeat each experiment three times and report the mean and standard deviation. AST-S model achieves 98.11$\pm$0.05, outperforms the SOTA model in~\cite{rybakov2020streaming}. In addition, we find AudioSet pretraining unnecessary for the speech command classification task as AST-S outperforms AST-P. 
To summarize, while the input audio length varies from 1 sec.~(Speech Commands), 5 sec.~(ESC-50) to 10 sec.~(AudioSet) and content varies from speech (Speech Commands) to non-speech (AudioSet and ESC-50), we use a fixed AST architecture for all three benchmarks and achieve SOTA results on all of them. This indicates the  potential for AST use as a generic audio classifier.

\section{Conclusions}

Over the last decade, CNNs have become a common model component for audio classification. In this work, we find CNNs are not indispensable, and introduce the Audio Spectrogram Transformer (AST), a convolution-free, purely attention-based model for audio classification which features a simple architecture and superior performance.


\section{Acknowledgements} This work is partly supported by Signify.

\newpage

\bibliographystyle{IEEEtran}

\bibliography{mybib}

\end{document}